\documentstyle[aps]{revtex}

\def\address#1{\begin{center}{\it #1}\end{center}}
\input psfig
\begin{document}
\newcommand{\bd}{\begin{displaymath}}
\newcommand{\ed}{\end{displaymath}}
\newcommand{\beq}{\begin{equation}}
\newcommand{\bea}{\begin{eqnarray}}
\newcommand{\eeq}{\end{equation}}
\newcommand{\eea}{\end{eqnarray}}
\newcommand{\la}{\label}
\newcommand{\r}{~\ref}

\def\R{{\bf R}}
\def\Z{{\bf Z}}
\def\del{\partial}
\def\Lap{\bigtriangleup}
\def\Div{{\rm div}\ }
\def\rot{{\rm rot}\ }
\def\curl{{\rm curl}\ }
\def\grad{{\rm grad}\ }
\def\Tr{{\rm Tr}}
\def\^{\omedge}
\def\goinf{\rightarrow\infty}
\def\goes{\rightarrow}
\def\bm{\boldmath}
\def\-{{-1}}
\def\inv{^{-1}}
\def\sqr{^{1/2}}
\def\isqr{^{-1/2}}
\def\oma{\!\!\!\!&=&\!\!\!\!}
\def\omb{\!\!\!\!&\equiv &\!\!\!\!}
\def\f{{\dot {\phi}}_0}

\def\reff#1{(\ref{#1})}
\def\vb#1{{\partial \over \partial #1}} 
\def\Del#1#2{{\partial #1 \over \partial #2}}
\def\Dell#1#2{{\partial^2 #1 \over \partial {#2}^2}}
\def\Dif#1#2{{d #1 \over d #2}}
\def\Lie#1{ {\cal L}_{#1} }
\def\diag#1{{\rm diag}(#1)}
\def\abs#1{\left | #1 \right |}
\def\rcp#1{{1\over #1}}
\def\paren#1{\left( #1 \right)}
\def\brace#1{\left\{ #1 \right\}}
\def\bra#1{\left[ #1 \right]}
\def\angl#1{\left\langle #1 \right\rangle}
\def\matrix#1#2#3#4#5#6#7#8#9{
        \left( \begin{array}{ccc}
                        #1 & #2 & #3 \\ #4 & #5 & #6 \\ #7 & #8 & #9
        \end{array}     \right) }
\def\manbo#1{}  

\def\d{{\rm d}}
\def\c#1{{}_{,#1}}
\def\tta#1#2{\theta^{#1}{}_{#2}}
\def\itta#1#2{\theta_{#1}{}^{#2}}

\def\hh{{h}}
\def\g{{g}}
\def\uh#1#2{\hh^{#1#2}}
\def\dh#1#2{\hh_{#1#2}}
\def\dg#1#2{\g_{#1#2}}

\newenvironment{proof}{{\bf Proof:}}{}
\newtheorem{theorem}{Theorem}[section]

\newcommand{\x}{x(r)} 
\newtheorem{lemma}[theorem]{Lemma}
\newtheorem{proposition}[theorem]{Proposition}
\newtheorem{corollary}[theorem]{Corollary}
\newtheorem{definition}{Definition}[section]
\newcommand{\be}{\begin{equation}}
\newcommand{\ee}{\end{equation}}
\newcommand{\ethbar}{\bar{\eth}}
\newcommand{\Ldbar}{\bar{\Ld}}
\newcommand{\p}{\partial} 
\newcommand{\omegabar}{\overline{\omega}}
\newcommand{\mbar}{\overline{m}}
\newcommand{\etabar}{\overline{\eta}}
\newcommand{\rhobar}{\overline{\rho}}
\newcommand{\sigmabar}{\overline{\sigma}}
\newcommand{\alphabar}{\overline{\alpha}}
\newcommand{\xibar}{\overline{\xi}}
\newcommand{\th}{\theta}
\newcommand{\Ld}{\Lambda\, }
\newcommand{\ld}{\lambda}
\newcommand{\tLd}{\bar{\Ld}\, }
\newcommand{\Wbar}{\overline{W}}
\newcommand{\debar}{\bar{\delta}}

\newcommand{\om}{\omega}
\newcommand{\ombar}{\bar{\omega}}
\newcommand{\Om}{\Omega}
\newcommand{\tOm}{\tilde\Omega}
\newcommand{\z}{\zeta}
\newcommand{\zb}{\bar{\z}}
\newcommand{\Ldp}{\partial_+\Ld}
\newcommand{\Ldm}{\partial_-\Ld}
\newcommand{\tLdo}{\partial_1\bar{\Ld}}
\newcommand{\tLdp}{\partial_+\bar{\Ld}}
\newtheorem{remark}{Remark}[section]

\newcommand{\Ldo}{\partial_1\Ld}
\newcommand{\Ldn}{\partial_0\Ld}
\newcommand{\varid}{\stackrel{\rm def}{=}}
\newcommand{\k}{\vec{k}}
\newcommand{\kcheck}{\check{k}}
\newcommand{\khat}{\hat{k}}
\newcommand{\mI}{(${m}_{I}$) }
\newcommand{\mII}{(${m}_{II}$) }
\newcommand{\scrip}{\cal J^+} 
\newcommand{\scripm}{\cal J^-}
\newcommand{\scri}{\cal J}
\newcommand{\ihat}{\hat{\imath}}
 \newcommand{\jhat}{\hat{\jmath}}
\newcommand{\Hspace}{${\cal H}$-space }
\newcommand{\lone}{\Ld_1} 
\newcommand{\lonebar}{\Ldbar_1}
\newcommand{\etatw}{\tilde \eta} 
\newcommand{\ztw}{\tilde \z}
\newcommand{\bareta}{\bar \eta}
\newcommand{\nn}{\nonumber}

\title{Null surfaces formulation   in 3D}

\author{ Diego M. Forni$^a$\footnote{E-mail: forni@fis.uncor.edu}
\and 
        Mirta Iriondo$^{a,b}$\footnote{E-mail: mirta@fis.uncor.edu}	
\and 
        Carlos N. Kozameh$^{a}$\footnote{E-mail: kozameh@fis.uncor.edu}}

\maketitle     
\address{${}^a$
        FaMAF,  
        Universidad Nacional de C\'ordoba, \\ 
        5000, C\'ordoba, Argentina
}
\address{${}^b$
Instituto de Investigaci\'on, Universidad Nacional de Villa Mar\'{\i}a,\\ V. Mar\'{\i}a, 5900, C\'ordoba, Argentina}

\begin{abstract}

The Null Surface Formulation of General Relativity is developed for 2+1
dimensional gravity.  The geometrical meaning of the metricity
condition is analyzed and two approaches to the derivation of the field
equations are presented. One method makes explicit use of the conformal
factor while the other only uses conformal information. The resulting
set of equations contain the same geometrical meaning as the 4-D
formulation without the technical complexities of the higher
dimensional analog.

A canonical family of null surfaces in this formulation, the light cone
cuts of null infinity, are constructed on asymptotically flat space
times and some of their kinematical aspects discussed. A particular
example, which nevertheless contains most of the generic features is
explicitly constructed and analyzed, revealing the behavior predicted
in the full theory.

\end{abstract}

\section{Introduction}
The Null Surfaces Formulation (NSF) developed by Frittelli, Kozameh, and Newman \cite{FKN1,FKN2,FKN3} gives 
an alternative treatment of Einstein's General Relativity (GR),
which emphasizes the dynamical role played in the theory by
the null hypersurfaces of the space-time metric. The construction is set up {\it{ab initio}} in terms of generic space-time foliations. These are represented by the
level surfaces of a two-parametric function $Z(x^a; \z,\zb)$ of the
$x^a$ coordinates, on which it is imposed the condition that the foliations become null for some space-time metric $\dg ab(x^a)$ to be obtained as a functional of $Z(x^a; \z,\zb)$. Then, an auxiliary function ${\Omega}(x^a; \z,\zb)$ fixes the resulting conformal freedom in such a way that the Einstein's field equations be satisfied for $\dg ab(x^a)$. Such equations are expressed entirely in terms of the non-local variables $Z(x^a; \z,\zb)$ and ${\Omega}(x^a; \z,\zb)$, with no mention at all of the metric and its associated tensors, which are merely functions of them. 

One important issue in the NSF is how to restrict the freedom of (infinitely many) different families of null hypersurfaces corresponding to the same metric. For asymptotically flat space-times a canonical family of null foliations can be selected by considering past null cones from points at future null infinity, $\scrip$. In this case the function $Z$ has a second meaning. For fixed values of $x^a$ the function $u = Z(x^a; \z,\zb)$ parametrically describes the intersection of the future light cone from $x^a$ with $\scrip$. This intersection is called a light cone cut of null infinity and it plays a central role in reformulating General Relativity in terms of the free data (representing outgoing gravitational radiation) at $\scrip$. 

It is worth mentioning that the final equations of NSF become a system of partial differential equations for the main variables that, in spite of the clear physical and geometric contents of the equations, are technically very involved and difficult to analyze. This is best exemplified in an alternative formulation  of NSF that decouples the original set of equations and produces a conformally invariant equation solely for $Z$, a desirable feature since $Z$ only contains information of the conformal structure \cite{Bach}. Although the resulting equation is equivalent to the vanishing of the Bach tensor, it cannot be explicitly written in terms of $Z$ and its derivatives on a reasonable number of lines.

Thus, from a purely technical point of view it is desirable to derive an analogous NSF formalism in a lower dimensional space. The obvious choice is 3-D gravity since its almost trivial character should simplify the formalism and make it easier to investigate all the hoped features that the formulation in four dimensions is required to posses in order to be consistent with standard GR. Examples of these are caustics and other singularities of null congruences predicted by general theorems, which are at present searched for only in a perturbative approach to the NSF field equations \cite{Bach}.

Moreover, three-dimensional GR can be used to study many examples of 4-D vacuum space-times with a space translation symmetry. By going to the ``manifolds of orbits'' of the associated Killing field, the problem is reduced to 3-D gravity coupled to a scalar field acting as matter source of the 3-D Einstein's equation. Assuming suitable fall off conditions on the scalar field, the concept of asymptotic flatness can be stated as close as
possible to the analogous four dimensional treatment \cite{Ash}. Consequently, for asymptotically flat 3-D space-times, one can introduce our canonical family of null surfaces, i.e., the light cone cuts of null infinity. In this way, we extend our formalism to 4-D space-times which are not asymptotically flat in the usual sense since they include cylindrical or axisymmetric waves.

Finally, from the quantum theory point of view, there is some evidence, at least at a linear level in 4-D \cite{fritelli} and at a non perturbative for some midi-superspaces in 3-D\cite{Tiglio}, that in contrast with the quantization in terms of the local fields like the metric, the quantization of the non-local variable $Z$ and ${\Omega}$ of NSF,  results in operators with non-diverging expectation values for semi-classical states. Concrete results have been obtained in 3-D since some midi-superspaces can be exactly solved and large quantum gravity effects appear for local fields such as the metric \cite{Ashtekar}. Recently, it has been shown that the quantum light cone cut admits a semiclassical approximation which is stable against small perturbations at transplanckian frequencies \cite{Tiglio}. Thus, it appears that this non local variable is best suited for semiclassical approximations. To develop a more thorough discussion on these issues one should have available the full classical theory.

This work is organized as follows. In section two we introduce one-parameter families of hypersurfaces in 3-D and discuss which conditions
should be imposed on these surfaces so that they become null. The main
results of this section, the metric components and metricity
conditions, were originally obtained by  E. Cartan and S. Chern  many
years ago \cite{Cartan,Chern} and recently rederived by M. Tanimoto
\cite{chino} but we follow an approach which is closely related to our
work done in four dimensions. In section three we derive the field
equations for these surfaces including matter sources.  As in previous
results one can decouple those equations and obtain a conformally
invariant equation for the null hypersurfaces. In section four we
review the notion of asymptotic flatness in 3-D and define the light
cone cuts. Several kinematical issues which differ from the four
dimensional case are pointed out and discussed. A particular example of
a light cone cut is constructed and analyzed. In the conclusions we
outline an approach to handle caustics and singularities that come
together with null surfaces.

\section{ NSF formalism in three dimensions.}

In this section the NSF formalism, as developed in \cite{FKN2}, is
adapted to dimension $2+1$ and the
corresponding kinematical structures are constructed. Although the
basic framework was outlined in detail in 4-D 
\cite{FKN2}, it is worth repeating the construction in 3-D since it is
conceptually equivalent to the higher dimensional case though
technically less involved and therefore easier to follow. 

We begin with a three-dimensional manifold $M$, and assume we are given one-parameter functions $Z(x^a, \z)$ of the space-time coordinates $x^a$; and parameter $\z$ on $S^1$ running between $0$ and $2\pi$. We also assume that for a fixed value of the parameter $\z$, the level surfaces of $Z$,
\begin{equation}
        Z(x^a, \z)={\rm const.},
        \la{surf}
\end{equation}
locally foliate the manifold $M$.

The statement that the level surfaces of $Z$ be indeed the characteristic or null surfaces of some metric $\dg ab(x^a)$ for $M$ is that for fixed values of $x^a$ and arbitrary values of $\z$ they satisfy
\beq   g^{ab}(x^a){\nabla}_{a} Z(x^a, \z){\nabla}_{b} Z(x^a, \z)=0.  \la{null}   
\eeq
An equivalent geometrical statement is that as the families of foliations intersect at a single but arbitrary point $x^a$ we demand that the enveloping surface forms the light cone from the point $x^a$. Thus, the parameter $\z$ spans the circle of null directions.

The idea now is to solve eq. (\ref{null}) for the five components of the conformal metric in terms of $\nabla_{a} Z(x^a, \z)$. Given an arbitrary function $Z$ the problem has no solution since we have an infinite number of algebraic equations (one for each value of $\z$) for five unknowns. Therefore we must impose conditions on $Z(x^a, \z)$ so that a solution exists. The solution and conditions are best expressed when written in a canonical coordinate system constructed from knowledge of $Z$.

To see this, we introduce three functions defined as 
\begin{equation}
(\theta^0,\theta^1,\theta^2)\equiv(u,\omega,r)\equiv(Z(x^a,\z),\del Z(x^a,\z),\del^2 Z(x^a,\z)),  \la{tita}
\end{equation}
where   ${\del}{\equiv}\frac{\del}{\del\z}$ denotes the derivative of $\z$ holding $x^a$ fixed. For each value of $\z$ these functions form a coordinate system intrinsically adapted to the surfaces. Its associated gradient $\tta ia$ and vector dual basis $\itta ia$ are given  by

\beq
{\tta ia}{\equiv}(d{\theta}^i)_{a},  
\quad {\itta ia}{\equiv}\left(\frac{\del}{\del{\theta}^{i}}\right)^{\!a}.
\eeq

We now demand that $u=const.$ is a null surface. Then on that surface $\om=const.$ singles out a null geodesic whereas $r=const.$ identifies a point on that geodesic. The next available scalar, $\del^3 Z(x^a, \z)$, determines the conformal metric and is constrained to satisfy the metricity conditions. Using eq. (\ref{tita}) and assuming the $\theta^i$ coordinate system is well behaved, one can obtain $x^a = x^a(\theta^i, \z)$ and define $\Ld(u,\om,r,\z) \equiv \del^3 Z(x^a(\theta^i, \z), \z)$.

A technical point worth mentioning is that the action of the operator $\del$, the parameter derivative holding $x^a$ fixed, acting on a function $F(u,\om,r,\z)$ is given by 
\beq
\del = \partial_\z + {\omega} \partial_u + r \partial_\om + \Ld \partial_r. \la{del}  \eeq
Thus, the action of the $\partial$ operator on $\Lambda$ is given by (\ref{del}).

Likewise, $\del$ does not commute with the directional derivatives $\partial_i{\equiv}\frac{\del}{\del\theta^i}$. The explicit form of the commutation relations, needed later, are

\be    
\la{comm1}  
[\p_u,\p]=\p_u\Ld \p_r,
\ee
\be    
\la{comm2}  
[\p_\om,\p]=\p_u+\p_\om\Ld \p_r,
\ee
\be    
\la{comm3}  
[\p_r,\p]=\p_\om +\p_r\Ld \p_r.
\ee

The metric components and metricity conditions are obtained by repeatedly operating $\partial$ on~($\!\!$\r{null}). Taking  $\partial$ on (\ref{null}) yields

\be \label{g01}
\partial(g^{ab}{\nabla}_{a} Z{\nabla}_{b} Z) = 2 g^{ab}{\nabla}_{a} Z {\nabla}_{b} \partial Z = 2 g^{01} = 0,
\ee
where we have used that $\partial g^{ab} = 0$. If we take again $\partial$ on the above equation we obtain

\be \label{g11}
g^{ab}(\nabla_{a} Z {\nabla}_{b} \partial^2 Z  +  \nabla_{a} \partial Z {\nabla}_{b} \partial Z)=  g^{02} + g^{11}= 0.
\ee
Repeating this procedure gives

\be \label{g12}
g^{ab}(\nabla_{a} Z {\nabla}_{b} \partial^3 Z  +  3 \nabla_{a} \partial Z {\nabla}_{b} \partial^2 Z)=  g^{02} \partial_r \Lambda + 3 g^{12}= 0,
\ee
\begin{eqnarray}
g^{ab}(\nabla_{a} Z {\nabla}_{b} \partial^4 Z  +  4 \nabla_{a} \partial Z {\nabla}_{b} \partial^3 Z + 3 \nabla_{a} \partial^2 Z {\nabla}_{b} \partial^2 Z) &=& 0, \label{M-1}\\
g^{02} \partial_r \partial \Lambda + 4 (g^{11}  \partial_{\omega} \Lambda + g^{12}  \partial_r \Lambda) + 3 g^{22} &=& 0.\nonumber
\end{eqnarray}
Using the commutator (\ref{comm3}), together with the already obtained metric components, we can rewrite this last expression as,

\be \label{g22}
g^{02} \left( \partial \partial_r \Lambda - 3 \partial_{\omega} \Lambda - \frac{1}{3} (\partial_r \Lambda)^2\right) + 3 g^{22} = 0.
\ee

Note that the four nontrivial metric components are proportional to $g^{02}$. It is therefore natural to take $g^{02}$ as a $\z$-dependent conformal factor $\Omega^2$ and write the metric tensor as $g^{ab} = \Omega^2 h^{ab} = \Omega^2  h^{ij}\theta^a_i\theta^b_j$ with 

\begin{equation}
        h^{ij}=
        \matrix 0010{-1}{-\frac{1}{3}\partial_r\Ld}1{-\frac{1}{3}\partial_r\Ld}
       {-\frac{1}{3}\del(\partial_r\Ld)+\frac{1}{9}(\partial_r\Ld)^2+\partial_\om\Ld}.
        \label{metric}
\end{equation}

The conformal factor $\Omega$ cannot be an arbitrary function of $\z$ since it is defined as
\[
\Omega^2 = g^{ab}\nabla_{a} Z {\nabla}_{b}\partial^2 Z.
\]
Therefore, 
\[
\partial \Omega^2 = g^{ab}( \nabla_{a}\partial Z  {\nabla}_{b}\partial^2 Z + \nabla_{a} Z  {\nabla}_{b}\partial^3 Z). 
\]
Thus, $\Omega$ and $\Lambda$ must satisfy
\begin{equation}
        3\del\Om =\Om \partial_r\Ld.
        \label{M1}
\end{equation}
Equation (\ref{M1}) is invariant under $\Om(x,\z) \to \Om'(x,\z) = f(x)\Om(x,\z)$
for an arbitrary $f(x)$. This freedom is a consequence of the conformal invariance of the formulation.

Note also that nothing prevents us from taking further $\partial$ derivatives on (\ref{M-1}). Since the r.h.s of this equation vanishes and all the components of the conformal metric have been  explicitly constructed from $\partial_i \Lambda$, the next $\partial$ on (\ref{M-1}) must impose a condition on $\Lambda$ so that a conformal metric exists. This equation can be written as

\begin{eqnarray}
g^{ab}(5\nabla_{a} \partial Z {\nabla}_{b} \partial^4 Z  + \nabla_{a} Z {\nabla}_{b} \partial^5 Z + 10 \nabla_{a} \partial^2 Z {\nabla}_{b} \partial^3 Z) &=& 0 \label{M2b}\\
5 g^{11} \partial_\om \partial \Lambda + 5 g^{12}  \partial_{r} \partial \Lambda + g^{02}  \partial_r \partial^2 \Lambda + 10 (g^{02} \partial_{u} \Lambda +  g^{12} \partial_\om \Lambda + g^{22}  \partial_{r} \Lambda) &=& 0.\nonumber
\end{eqnarray}

Using the explicit form of the metric components and the commutation relations and dividing out by the conformal factor gives
\begin{equation}
       M[\Lambda] \equiv 2\left( \del(\partial_r\Ld) -\partial_\om\Ld -{2\over9}(\partial_r\Ld)^2\right) \partial_r\Ld
        -\del^2(\partial_r\Ld)+3\del(\partial_\om\Ld)-6\partial_u\Ld=0.
        \label{M2}
\end{equation}
Conversely, if (\ref{M2}) is satisfied, then further $\partial$ derivatives on (\ref{g22}) are identically satisfied.

We summarize this section as follows: given a function $Z(x^a,\z)$ such that $\Lambda(\theta^i,\z) \equiv \partial^3 Z(x^a,\z)$ satisfies $M[\Lambda] = 0$, then $Z = const.$ is a null hypersurface of $g^{ab}$.

{\bf Remark 1:} Equation (\ref{M2}) is the main metricity condition, (\ref{M1}) is only used to fix the $\z$-dependence of the conformal factor. In fact, given a function $Z(x^a,\z)$ such that $\partial^3 Z(x^a,\z)$ satisfies $M[\Lambda] = 0$, then $Z(x^a,\z) = const.$ is a null hypersurface of $h^{ab}(x^a,\z')$ for an arbitrary value of $\z'$.

To see this we first show that $h^{ab}(x^a,\z')$ satisfies
\begin{equation}
        \del h^{ab} =-\frac{2}{3} \partial_r\Ld h^{ab},
        \label{M3}
\end{equation}
which follows from the definition of $h^{ab}$. The above equation can be integrated immediately to give

\[
h^{ab}(x^a,\z') = \mbox{ exp} \left(-\frac{2}{3} \int^{\z'}_{\z} \partial_r\Ld (x^a,\eta) d\eta \right) h^{ab}(x^a,\z).
\]
Thus, if $Z(x^a,\z)$ yields null surfaces for $h^{ab}(x^a,\z)$ it also gives null surfaces for any $h^{ab}(x^a,\z')$ that satisfies (\ref{M3}). Since  $h^{ab}(x^a,\z)$ is an explicit functional of $\Lambda$ we conclude that (\ref{M2}) is the only condition to be imposed on $Z$  so that a conformal  geometry can be constructed on the three dimensional manifold.

{\bf Remark 2:} In fact, it is not necessary to assume that $ g^{ab} = g^{ab}(x)$ to obtain the explicit form of the conformal metric and metricity condition  (\ref{M2}). If the metric $g^{ab}$ is such that 
\[
\partial g^{ab} = \mu (x, \z) g^{ab}
\]
for an arbitrary function $\mu (x, \z)$, we can repeat the steps shown above to obtain the same conformal metric $h^{ab}(x^a,\z)$ and metricity condition (\ref{M2}).

{\bf Remark 3:} Starting from a completely different perspective, E. Cartan \cite{Cartan} and S. Chern \cite{Chern} studied the geometry of the                                                                                                                                  ordinary differential equation
\[
\partial^3 Z = \Lambda(\z, Z, \partial Z, \partial^2 Z).
\]
They found that the condition (\ref{M2}) has a simple geometrical meaning, namely, that the Cartan normal conformal connection defined in the solution space is torsion free\cite{Chern}. It is therefore natural in our case to obtain (\ref{M2}) as a metricity condition.

{\bf Remark 4:} A very interesting (and open) problem in 4-D is to understand the geometrical meaning of the metricity conditions\cite{FKN1}. It is natural to conjecture that the complex PDE derived in that case is also the torsion free condition for the conformal connection defined in the solution space of the starting second order PDE for $Z$ on the sphere of null directions.

\section{ The field equations.}

In this section we derive the field equations for the basic variables of this formalism. In contrast with the 4-D case, the vanishing of the trace free part of the Ricci tensor does not contain interesting solutions. Since the Weyl tensor vanishes in three dimensions, the solutions of 
$$R_{ab}- \frac{1}{3} R g_{ab}=0$$
yield spaces with constant curvature. Thus, we must add a stress energy tensor as a source term on the r.h.s. of the above equation. For simplicity we will assume that $T_{ab}$ is a given trace free tensor. If the matter source contains a trace part then the trace equation is automatically satisfied by virtue of the vanishing of the divergence of $T_{ab}$. This can be easily seen by writing

\[
G_{ab}  = \underline{G}_{ab} + \frac{1}{3} g_{ab} G,
\]
where $G$ and $\underline{G}_{ab}$ are the trace and the trace free part of the Einstein tensor respectively. If the trace free equations
\[
\underline{G}_{ab}  = \underline{T}_{ab},
\]
are satisfied, then using the vanishing of the divergence of the stress energy tensor and the Einstein tensor one obtains 
\[
\nabla_a(G-T) = -4 \nabla^b(\underline{G}_{ab}- \underline{T}_{ab}) = 0.
\]

We thus start with the Einstein equations and contract the indices $a,b$ with the null vectors $\itta 2a \itta 2b$. Note that this operation automatically removes the trace part of the equation.  Rewriting the resulting equation in terms of our variables yields 

\be
\label{Omega}
\partial^2_r\Omega=T_{rr}\Omega.
\ee
At a first glance it appears that this single equation cannot be equivalent to the five components of the Einstein equations. Note, however, that (\ref{Omega}) is true for any value of $\z$. Thus, if we add to (\ref{Omega}) the metricity conditions which gives the change of $\Om$ and $\Ld$ under variations of $\z$ we obtain a consistent set of equations equivalent to the standard Einstein equations. The final equations read,
\be
\label{eq:system}
\left .\begin{array}{ll}
(\mbox{E})\qquad\partial^2_r\Omega&=T_{rr}\Omega,

\vspace{2mm}

\\

\vspace{2mm}

(m_{I})\qquad\del\Omega&=\displaystyle{{1\over3}}\partial_r\Ld\Omega,

\vspace{2mm}

\\

( m_{II})\qquad 6\partial_u\Ld&=-{4\over9}(\partial_r\Ld)^3+2\partial_r\Ld\del(\partial_r\Ld)
        -2\partial_\om\Ld\partial_r\Ld
        -\del^2(\partial_r\Ld)+3\del(\partial_\om\Ld). 
\end{array}\right \}
\ee

The above constitute a coupled set of equations for $\Om$ and $\Ld$. These two functions, however, have completely different meaning; whereas $\Ld$ determines the entire conformal structure, $\Om$ only fixes the scale. It is thus desirable to decouple the above set of equations and obtain equations that only involve $\Ld$. By construction, the resulting equations would be conformally invariant.

The approach followed to decouple the field equations is to study the integrability conditions for \mI and (E) since both equations must be satisfied by a  single function $\Omega $. Therefore, conditions are imposed on $\Ld$ so that a solution exists.

To illustrate the procedure and to discuss the freedom left in the solutions we first consider the case when $T_{rr}=0$. Using (\ref{comm1}), (\ref{comm2}) and (\ref{comm3}),   we calculate  
\be
\label{com11}
[\p^2_r,\del]\Omega=2\p_\om\p_r\Omega+2\p_r\Ld\p^2_r\Omega+\p^2_r\Ld\p_r\Omega.
\ee
On the other hand,  using \mI and (E), we have 
\be
\label{com12}
[\p^2_r,\del]\Omega=\p^2_r({1\over3}\p_r\Ld\Omega),
\ee
the r.h.s. of equations (\ref{com11}) and (\ref{com12}), gives
$$
2\p_\om\p_r\Omega={1\over3}\p^3_r\Ld\Omega-{1\over3}\p^2_r\Ld\p_r\Omega.
$$
Applying $ \p_r$ to the above equation, commuting the partial derivatives, and using eq. (E) yields
\be
\la{B212}
\p^4_r\Ld=0.
\ee

Thus, for spaces with constant curvature the field equations for $\Ld$  are,

\begin{eqnarray}
\p^4_r\Ld&=&0, \la{B}\\
6\p_u\Ld&=&-{4\over9}(\partial_r\Ld)^3+2\partial_r\Ld\del(\partial_r\Ld)
        -2\p_\om\Ld\partial_r\Ld
        -\del^2(\partial_r\Ld)+3\del(\partial_\om\Ld). \la{C}
\end{eqnarray}
By construction, equation (\ref{B}) is conformally invariant. It is thus interesting to ask what tensorial quantity is represented by $\p^4_r\Ld$. It can be easily shown that
$$
\frac{1}{6}\p^4_r\Ld = B_{ab}Z^aZ^b,
$$
with
$$
B_{ab} = 2\epsilon_a{}^{mn}\nabla_{[m}S_{n]b},
$$
$$
S_{ab} = R_{ab} -\frac{1}{4}R g_{ab}.
$$

$2\nabla_{[m}S_{n]b}$ known as the Bach tensor in the literature, was first defined and used by E. Cotton\cite{Cotton} in 1899 to show that its vanishing determines the conformal structure of spaces with constant curvature (see Appendix A).

The solutions to equations (\ref{B}) and (\ref{C}) should contain functional degrees of freedom since any null surface compatible with a given Einstein metric should be a solution of those equations. This feature is shown below where we discuss the simplest set of equations corresponding to flat space and we obtain the collection of all null hypersurfaces in Minkowski space. For this we consider the special class of solutions to the above equations given by
$\Ld$'s that satisfy

$$\p_r\Ld = 0, \;\;\; \p_\om\Ld = -1,$$
since this restriction yields a Minkowski metric (see eq. (\ref{metric})). We immediately see that equation (\ref{B}) is identically satisfied whereas the other equation gives
$$\p_u\Ld = 0.$$
Thus, the solution can be written as
$$\Ld + \om = g(\z), $$
where $g$ is an arbitrary function on $S^1$. Using the relationship between $\Ld$, $\om$, and $Z$ we rewrite the above equation as
$$\partial^3 Z + \partial Z= g(\z),$$
whose solution contains three arbitrary constants. Denoting those constants as $(t,x,y)$ we write the solution as
$$ Z= t + x \cos(\z) + y \sin(\z) + \alpha(\z),$$
with $\partial^3 \alpha = g(\z)$. The above solution is the generating function for an arbitrary null hypersurface in flat 3-D space\cite{eikonal}. Its freedom is given by an arbitrary function on the circle of null directions. In the next section we will cut down this freedom by finding a canonical family of surfaces.

Following a similar approach we obtain the field equations for $\Ld$ when matter source is included. Then eq. (\ref{B}) is replaced by (see Appendix)

\be \label{bach2}
\p^4_r\Ld =  6 \p_\om T_{rr}- 6\p_r T_{r\om}+\p_r^2 \Ld + 6\Om^{-1} (T_{rr} \p_\om \Om - T_{r\om} \p_r \Om  ).
\ee                                          

Note that $\Omega$ appears explicitly in the above equation. Thus, it would appear that we are unable to get rid of the conformal factor. This is  not so. Applying several times the $\p$ operator on (\ref{bach2}) gives five independent equations involving the three components of $\nabla_a \Om$. We can algebraically solve for these three components and leave two extra equations for $\Ld$ and the energy-momentum tensor. An explicit illustration of this procedure written in tensorial language is given in Appendix B.

\section{Light Cone Cuts of $\scrip$ }


\indent
The main results obtained in previous sections, the metricity conditions and the field equations, are valid for any family of null hypersurfaces. In this section we introduce a canonical family with a well defined geometrical meaning. They represent past null cones from future null infinity. As a necessary step before introducing this family of null hypersurfaces we first review the notion of asymptotic structure available in 3-D\cite{Ash}.

A space-time is considered to be asymptotically flat at null infinity if one can attach a boundary $\scrip$  with topology $S^1\times R$ to the original manifold, a smooth metric field $g'_{ab}$ and scalar field $\om$ on $M\cup\scrip$, such that:
\begin{enumerate}
\item on $M$ $\om \neq 0$, $g'_{ab}= \om^2 g_{ab}$,
\item on $\scrip$ $\om = 0$, $\om_a = \nabla_a \om \neq 0$, and $ g'^{ab} \om_a \om_b = 0$,
\item if $\om$ is chosen such that $\nabla'^a \om_a=0$ on $\scrip$, then the vector field $n^a = g'^{ab} \om_b$ is complete on $\scrip$.
\end{enumerate}

As in 4-D, the smooth unphysical metric on $M\cup\scrip$ is used to discuss the fall off behavior of matter fields near $\scrip$. By bringing null infinity at a finite distance, one can use the techniques of differential geometry near 
$\scrip$ instead of taking limits along null directions. Moreover, the space of null geodesics is the same for both the physical and unphysical metric, the only difference being in the corresponding affine parametrization of the geodesics.

There are, however, several differences with 4-D asymptotia. In 3-D the Weyl tensor vanishes and the metric is flat outside sources. Thus, the dynamical degrees of freedom are coded in the behavior of matter fields near $\scrip$.

We now turn our attention to the description of light cone cuts of $\scrip$
for 3-D asymptotically flat space-times. A straightforward method to
obtain these cuts is to start with the future light cone from an arbitrary
point $x^a$ of the space time and define {\it a  light cone cut of null infinity} as the intersection of this
cone with the null boundary. Using Bondi coordinates $(u,\z)$ at
$\scrip$ one can locally describe this intersection as
\be \la{cuts}
u = Z(x^a,\z).
\ee
The function $Z$ is called the light cone cut function. It
is easy to show that $Z$ has a second meaning. For fixed values of $u$
and $\z$, the points $x^a$ that satisfy (\ref{cuts}) lie on the past
null cone from the point $(u,\z)$ at $\scrip$. Thus, the level surfaces of
$Z(x^a,\z)$ form a canonical one-parameter family of null surfaces on the space time.

In the above construction one makes explicit use of the background
metric to first obtain the null cone from $x^a$ and then intersect this
cone with $\scrip$. Since our goal is to regard the cuts as the main
variables, we search for a construction that does not involve
explicit use of the metric. The idea is to define the cuts as the
solutions to a third order ODE where the source term, $\Lambda$, is obtained
from field equations that satisfy the desired asymptotic conditions.

As was mentioned in the previous section one has available two sets of
field equations for $\Ld$ and it is worth reviewing both approaches
since they involve different variables. The first set of equations,
namely eq. (\ref{eq:system}), are PDE's for $\Omega$ and $\Lambda$ with
the matter field $T_{rr}$ as a source term. Imposing regularity and
asympotic conditions on the stress-energy tensor as discussed in
\cite{Ash} together with apropriate initial conditions on  $\Omega$ and
$\Lambda$ one should be able to find solutions that preserve the
asymptotic structure. For example,  $\Omega$ and $\p_\om\Ld$ should tend
to 1 and -1 respectively as we approach to $\scrip$. Assuming one has
solutions to  (\ref{eq:system}) and denoting those solutions as

$$
\Omega (u, \om, r, \z)  = \Omega [T_{rr}],\\
\Lambda (u, \om, r, \z) = \Lambda [T_{rr}],
$$
where the term in bracket denotes functional dependence, one should then use the obtained $\Ld$ to define the light cone cuts as the solutions to 

\be \la{field2}
\partial^3 Z = \Ld(Z, \p Z, \p^2 Z, \z) = \Lambda[T_{rr}],
\ee
where the coordinates $(u,\om,r,\z)$ have been replaced by $(Z,\p Z,\p^2 Z,\z)$.
The resulting equation is a third order ODE with a 3-dim solution space. Note that the metricity condition is automatically satisfied since it is part of the field equations for $\Ld$. Thus, one can algebraically obtain a metric in the solution space such that the level surfaces $Z= const.$ are null surfaces of that metric.

If we follow the second approach we should start with eqs.(\ref{C}) and (\ref{bach2}) and follow a similar construction to end up with (\ref{field2}). The implicit assumption being done here is that one is able to eliminate $\Omega$ in (\ref{bach2}) and obtain a set of field equations for $\Ld$. Although this second approach offers the advantage of having just one variable the first set of equations for $\Omega$ and $\Lambda$ appears to be technically more tractable and easier to handle.

At this point we mention a possible third approach that yields a different equation for the light cone cuts. We first derive an analog of the
Sachs' theorem in four dimensions (which gives a relationship between
the cuts and the shears associated with the Bondi and light cone cuts
\cite{cortes}). Since the shear vanishes in three dimensions we use the only
available structure that is left, the divergence of null congruences.
To be more precise, we start with the divergence $\rho$ of a null cone
with apex at $x^a$. This optical scalar satisfies the following ODE
\be \la{divergence}
\frac{\partial \rho}{\partial s} = \rho^2 + \Phi_{oo},
\ee
where s is an affine length and $\Phi_{oo} = R_{ab}\l^a\l^b$. 
The null cone congruence condition is imposed by assuming that near the apex 
$s=s_o$, $\displaystyle{\rho = \frac{-1}{s-s_o}}$. Following a similar 
calculation outlined in \cite{Bach} one can show that at points of 
intersection between the light cone from $x^a$ and $\scrip$, the Bondi and 
light cone divergences are related through

\be \la{field1}
\partial^2 Z = \rho_B - \rho_{Z},
\ee
where $\rho_B$ is the divergence of the Bondi cuts (usually taken to be zero) and $\rho_{Z}$ the solution of (\ref{divergence}) evaluated at $\scrip$.
Note that $\rho_{Z}$ is a functional of the Ricci tensor $\Phi_{oo}$. Thus, we can regard (\ref{divergence}) and (\ref{field1}) as the field equations for $Z$.
Note also that the resulting equation is a second order ODE which in principle is equivalent to the third order ODE previously obtained. However, the solution space of this last equation is 2-dim and it is not clear at this point how to generate the required extra dimension. A thorough discussion of these issues will be presented in the future.

A very interesting and difficult problem is to analyze the behaviour of the solutions to the field equations. Since these solutions represent characteristic surfaces, they will have caustics and self-intersections. It can be shown that both $r=\p^2 Z$ and $\Ld=\p^3 Z$ diverge at conjugate points. Thus, the field equations as written above are not suited to analyze the behaviour of the solutions in the neighborhood of conjugate points. Possible alternative formulations are discussed in the last section.

\subsection{Light cone cuts: an example}

In this subsection we find the light cone cuts associated with particular simple models of space-times. These cuts of null infinity exhibit the typical behaviour that one has to deal with in NSF, namely, the presence of caustics and self-intersections.

As we show below, the simplest axisymmetric and static line element,
\be \label{m1}
{{\it ds}}^{2}=-d{t}^{2}+{\rho}^{-8\,M}{\!}\left(d{\rho}^{2}+{\rho}^{2}d{\phi}^{2
}\right)
 \eeq
with a constant $M$ and polar coordinates $\left(t,\rho,{\phi}\right)$ does not give interesting cuts of $\scrip$. Nevertheless it is useful to follow the construction since it gives us a hint of how to modify the background space time to obtain non trivial cuts. 

The metric (\ref{m1}) is interpreted as the space-time of a particle at the origin since its Einstein tensor $G_{ab}$ is given by $M{\delta}(\rho)\nabla_a t \nabla_b t$\cite{deser}.

After the coordinate change $r={\frac {{\rho}^{\alpha}}{\alpha}}$ with ${\alpha}=1-4 M$, (no relationship with the canonical coordinate $r$ defined in previous sections) it can be put in the form 
 \beq   {{\it ds}}^{2}=-d{t}^{2}+d{r}^{2}+{\alpha}^{2}{r}^{2}d{\phi}^{2},   \la{family}     \eeq
which is a locally flat metric with a global deficit angle of $2{\pi}\left(1-{\alpha}\right)$. Since the solution is not regular at the world-line of the particle $\rho\!=\!0$, this line must be removed from the otherwise flat space-time. Thus, the integration of the geodesic equation is trivial except for the following. For any apex $x^a$ there will be one null geodesic that will point to the origin and will not reach $\scrip$. Therefore, the light-cone cuts at null infinity will be open curves without caustics.

In order to obtain closed cuts of null infinity, we must then search
for a matter source with a non-singular metric. The simplest choice is
a ring of mass whose metric can be thought of a flat interior matched
to a vacuum external solution of the form~($\!\!$\r{family})\cite{Ash}:

\be
ds ^2 = \left\{\begin{array} {ll}
 -dt^2 + dr^2 + r^2 d\phi^2 , \;\;\;\;\;\;&  0 \leq r \leq  R \;,\\
 -dt^2 + dr^2 + \displaystyle{\left(\frac{ R}{ R + a}\right)^2}({r + a})^2 d\phi^2  , \;\;\;\;\;\;&   R \leq r,  \la{ring} 
\end{array}
\right .
\ee
where the parameter $ R$ is the radius of the ring and ${a}$ is related to the mass. A calculation of its energy momentum gives\cite{tutan},

\beq  T_{ab}=\frac{a}{8{\pi} R( R+a)}{\delta}(r- R)\nabla_a t \nabla_b t.  \eeq
Integration on a spacelike surface gives $\displaystyle{M=\frac{a}{4( R+a)}}$, so that $M \geq 0$ implies that we must take $a \geq 0$. The Minkowski space corresponds to $a=0$. There is another range of negative $a$'s that also gives a positive mass but these paramenters are not continuously connected to  $a=0$.

To solve the null geodesics equation and obtain the null cones we introduce the  retarded time coordinate $u\!=\!t-r$, and rewrite (\ref{ring}) as
\be \label{metric2}
ds ^2  = -du^2 - 2du\,dr + {\x}^2 d\phi^2,  
\eeq
with
\be
\x = \left\{\begin{array} {ll}
 r, \;\;\;&  0 \leq r \leq  R,\\
 R\displaystyle{\frac{(r+a)}{( R+a)}}.\;\;\;&   R \leq r \label{ring2}
\end{array}
\right .
\ee

A first integral of the geodesic equations can be obtained by observing that $k^a k_a = 0$ and that the scalar product between the null vector $k^a$ and the Killing fields $\chi_u^a$ and $\xi_\phi^a$ are constant along the null geodesic. 

 \begin{eqnarray} 
k_a\chi_u^a &=& -{\dot u}-{\dot r} = -E, \la{1}\\
k_a\xi_\phi^a &=&  \x^2{\dot{\phi}} = L,\la{2}\\
k_ak^a &=& -{\dot u}^2-2{\dot u}{\dot{r}}+\x^2{{\dot {\phi}}}^2 = 0.  \la{3}
 \end{eqnarray}
 Then, inserting equations (\r{1}) and ~($\!\!$\r{2}) in~($\!\!$\r{3}) gives

 \beq
    {\dot r} = {\pm} E \sqrt{1-\left({\frac{b}{\x}}\right)^2} ;  \la{rpto}   
\eeq
with $b\equiv L/E$, and from ~($\!\!$\r{1}), ~($\!\!$\r{2}) and ~($\!\!$\r{rpto}), we get
\bea  {\frac {d{u}}{d{r}}} &=& {\pm} \frac {1{\mp}\sqrt{1-\left({\frac{b}{\x}}\right)^2}}{\sqrt{1-\left({\frac{b}{\x}}\right)^2}}, \la{du/dr} \\
   \frac{d{\phi}}{dr} &=& {\pm} \frac{b}{{\x}^2 \sqrt{1-\left({\frac{b}{\x}}\right)^2}}.\la{dphi/dr}
\eea

Integrating (\r{du/dr}) and (\r{dphi/dr}) between $r\!=\!r_0$ and $r\!=\!{\infty}$, gives the desired light cone cut. The more interesting case occurs when $r_0 > R$ so we will restrict ourselves to this case. We now consider the circle of null directions from the point $(u_0, r_0, \phi_0 = 0)$ and divide this circle in four quadrants. Since the solution has axial symmetry with respect to the line that joins $r_0$ with the origin we will only consider the upper quadrants which are distinguished by the initial value of ${\dot r}(r_0)$. The integration on the quadrant where ${\dot r}(r_0)\geq0$ is straightforward and does not give any caustics. (Along those lines one takes the positive sign in (\r{du/dr}) and (\r{dphi/dr}).) The value of $b$ determines the starting null direction. This parameter ranges between $0$, when it points away from the origin and $b_{m} = x(r_0)$, when  ${\dot r}(r_0)\!=\!0$. It is does appropriate to introduce a null angle $\displaystyle{0\leq \theta \leq \frac{\pi}{2}}$ defined by $\displaystyle{\sin \theta = \frac{b}{b_m}}$.

Care must be taken in the quadrant where ${\dot r}(r_0)\leq0$. Here $b$ also ranges between $b_{m}$, and $0$, when it points to the origin. Then, the range of $\theta$ in this  quadrant is given by  $\displaystyle{\frac{\pi}{2}\leq \theta \leq \pi}$.  However, there exists a critical value of b, called $b_{c}$  (and a corresponding $\theta_c$) such that for $b\leq b_c$ the geodesics enter the ring. This value of $b_{c} =  R$ is obtained from the condition that ${\dot r}( R)=0$. 

We start the integration with ${\dot r} < 0$ until the turning point ${\dot r}(r_t)\!=\!0$ is reached. At this point $x(r_t) = b$ from which the value of $r_t$ is obtained. Then, for  $r_t \leq r \leq \infty$ we choose the positive  sign in  (\r{du/dr}) and (\r{dphi/dr}) to finish the integration.

The light cone cut of ${\cal J}^+$ from a point with coordinates $({u}_{0}, {r}_0, {\phi}_{0}=0)$ will be given by the following $\theta$-parametrized relations between $u$ and ${\phi}$:
 \bea  
u-u_{0} &=& (r_0 +a)(1-\cos \theta), \\
{\phi} &=& \theta\left(1+\frac{a}{R}\right), \qquad 0 \leq \theta \leq \theta_c
 \eea

 \bea  
u-u_{0} &=& (r_0 +a)(1-\cos \theta)
 - 2a\sqrt{1-\left(\frac{r_0+a}{R+a}\sin \theta\right)^2} , \la{crossu}\\
{\phi} &=& \theta\left(1+\frac{a}{R}\right)- 2\frac{a}{R} \arccos(\frac{r_0+a}{R+a} \sin \theta), \qquad \theta_c \leq \theta \leq \pi  \la{crossp}  \; ;   
\eea

We now give a brief analysis of the singularity structure of the light cone cuts for this particular case. Without involving heavy use of singularity theorems we can see that caustics will occur when $\phi$ is not an injective function of $\theta$. (If otherwise, we could invert the relation to obtain $\theta = \theta(\phi)$ insert this back into $u=U( \theta(\phi))$ and trivially show that the cut is the graph of a function without self intersections.)

The graphs $\phi$ vs $\theta$ for fixed values of $R=3$ and $a=1$ and two values of $r_0 = 5, 10$ given below serve as illustrations of the most general situation.
 \begin{figure}
\centering
{\hspace*{-.1in}\psfig{file=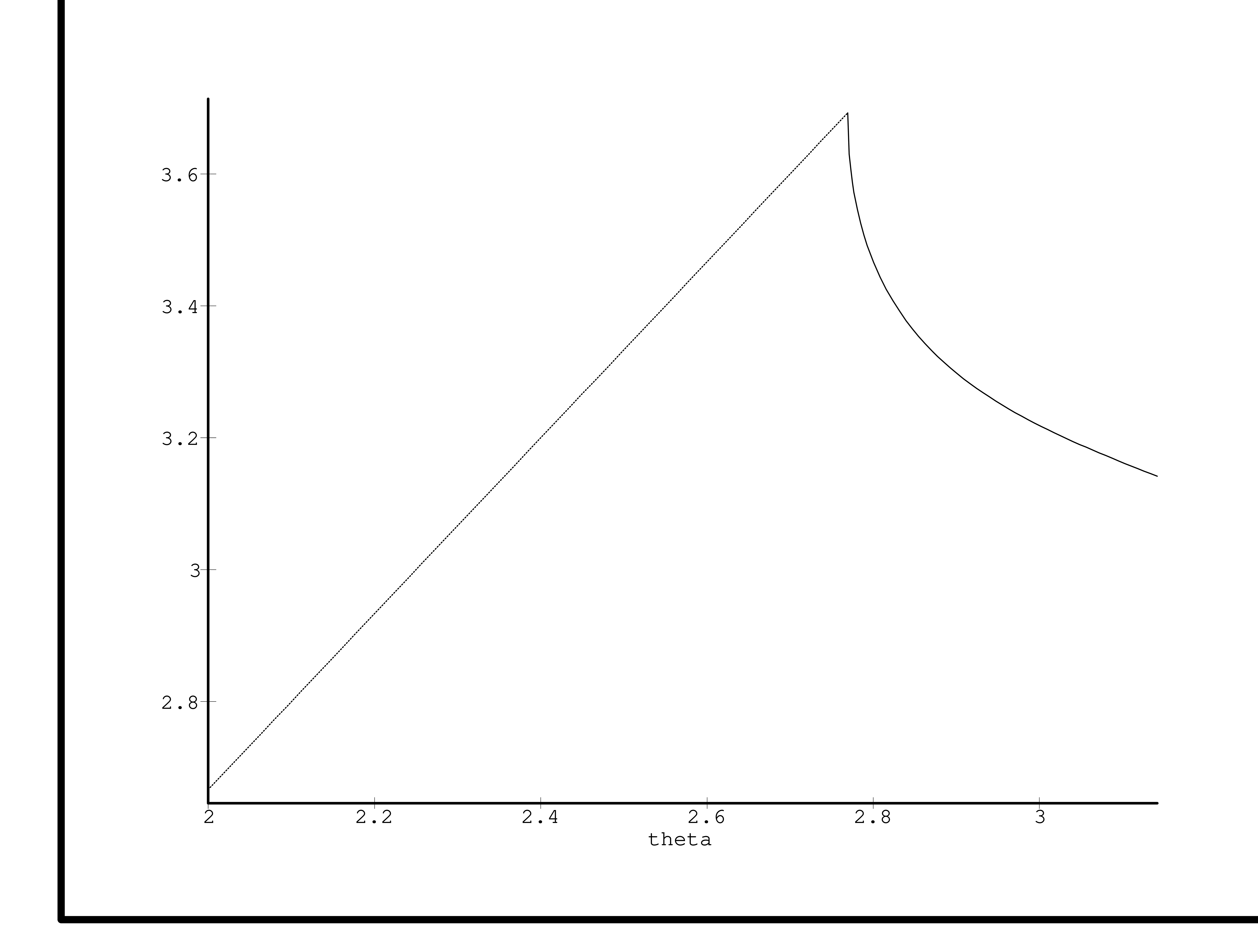,height=3in,width=2.5in}}\qquad
\mbox{{\hspace*{-.1in}\psfig{file=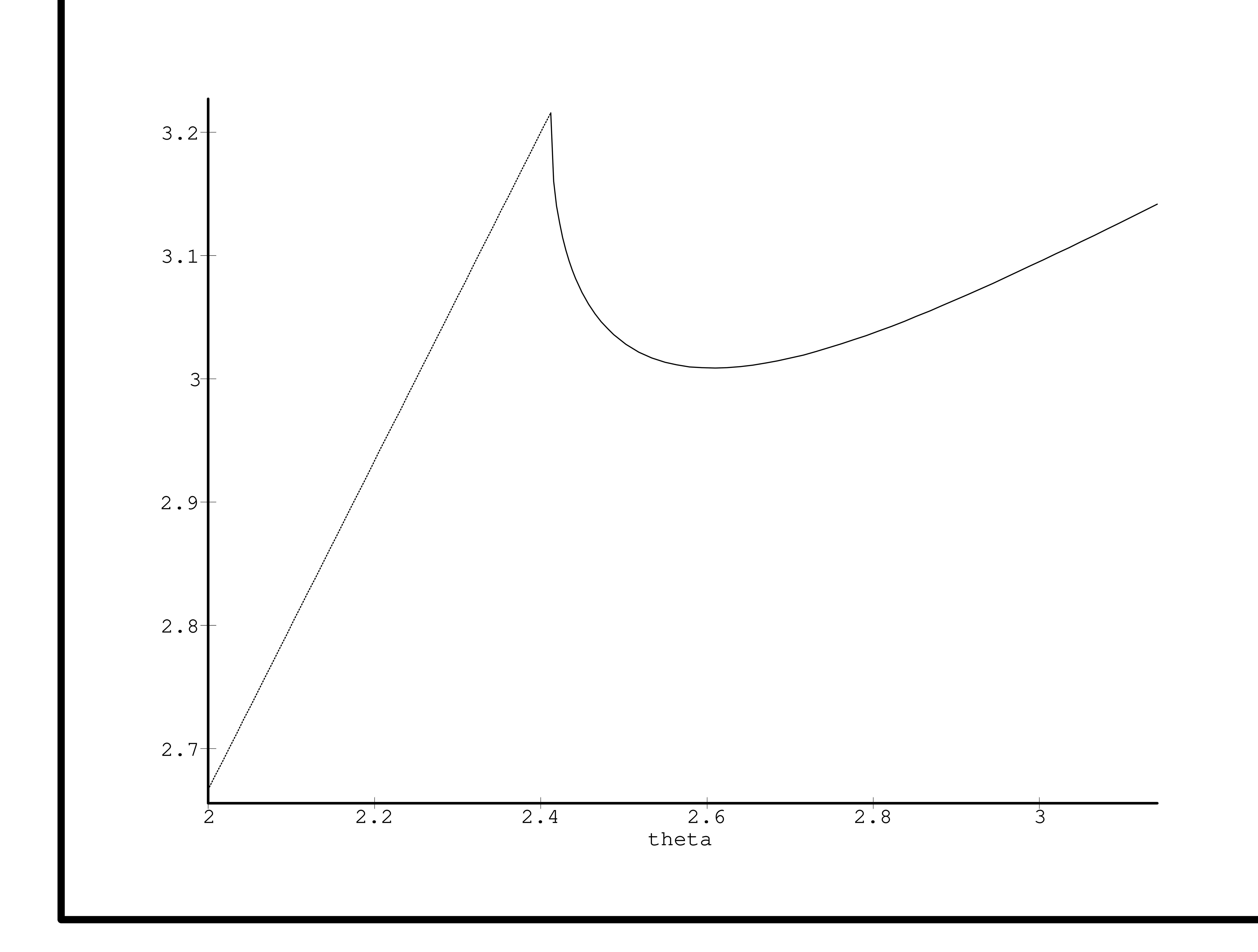,height=3in,width=2.5in}}}
\caption{ $\phi$ vs $\theta$ for (a) $r_0=10$ and (b) $r_0=5.$ }
\end{figure}

In Fig. 1.a we observe that around $\theta = \theta_c$ the function fails to be injective. This case, however, is not analyzed in the theory of caustics since at $\theta_c$ the function is not differentiable. We thus adopt the criteria that whenever the function $\phi(\theta)$ is continuous but not differentiable, we define a caustic point to be such that there is a change of sign of the left and right derivatives at this point. It is worth mentioning that the non differentiability around this caustic point follows from the fact that the Christoffel symbols, the coefficients of the geodesic equation, have a jump at $r=R$. By smoothing out the shell with a mass density function we can recover differentiability in the solution of the geodesic equation.

For the other value $r_0 = 5$ (see Fig. 1.b), we note the presence of a second caustic point $\theta_0$ where $\displaystyle{\frac{d\phi}{d\theta}}(\theta_0)=0$. 

In general, we will always have a caustic point $\theta_c$ for any value of $R$, $a$, and $r_0$ and a second caustic point whenever $r_0$ ($>R$) satisfies the condition

\[
2 a (r_0+a) < (R+a)^2.
\]
We can also plot the corresponding light cone cuts. 
 \begin{figure}
\centering
{\hspace*{-.15in}\psfig{file=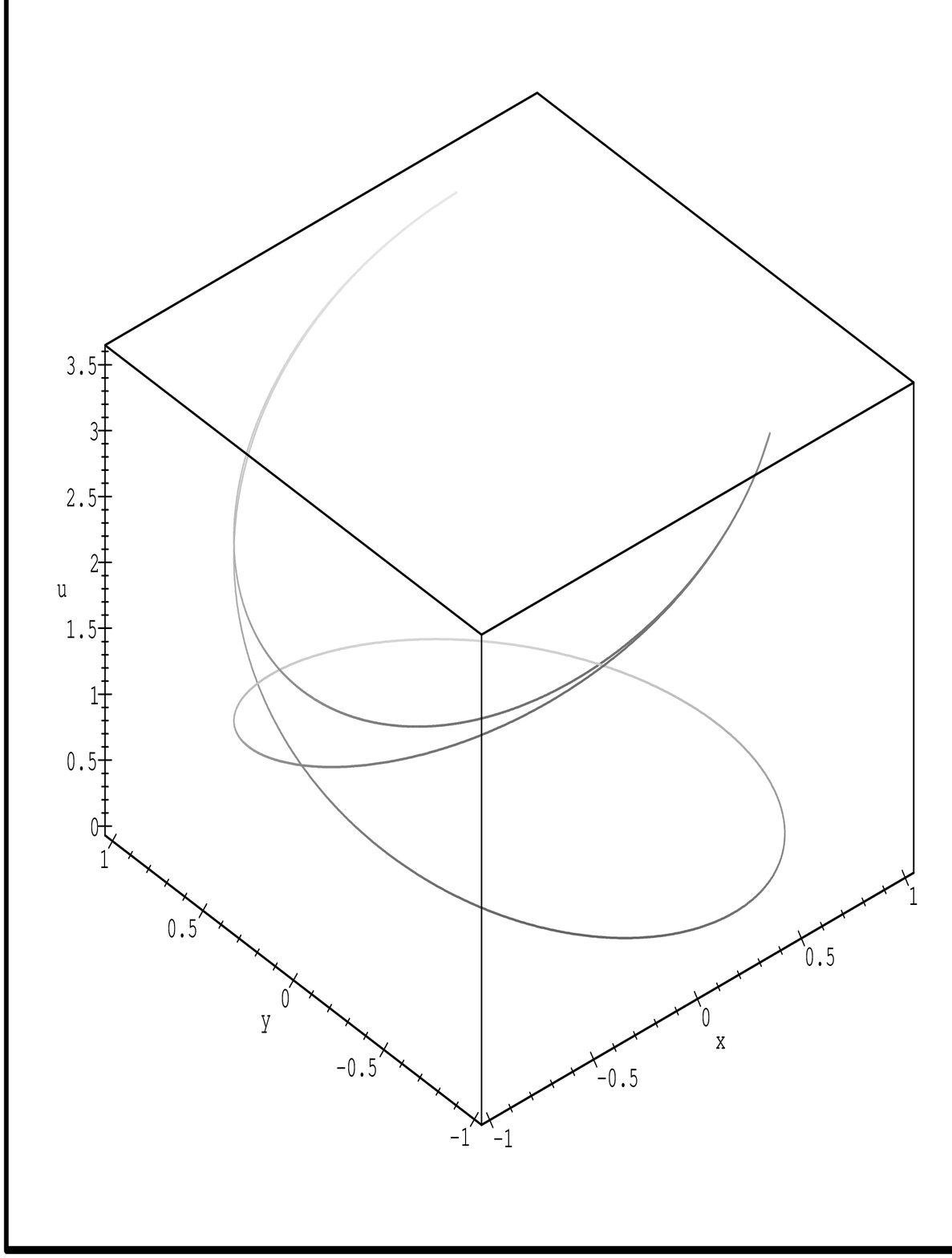,height=3in,width=2.5in}}\qquad
\mbox{{\hspace*{-.15in}\psfig{file=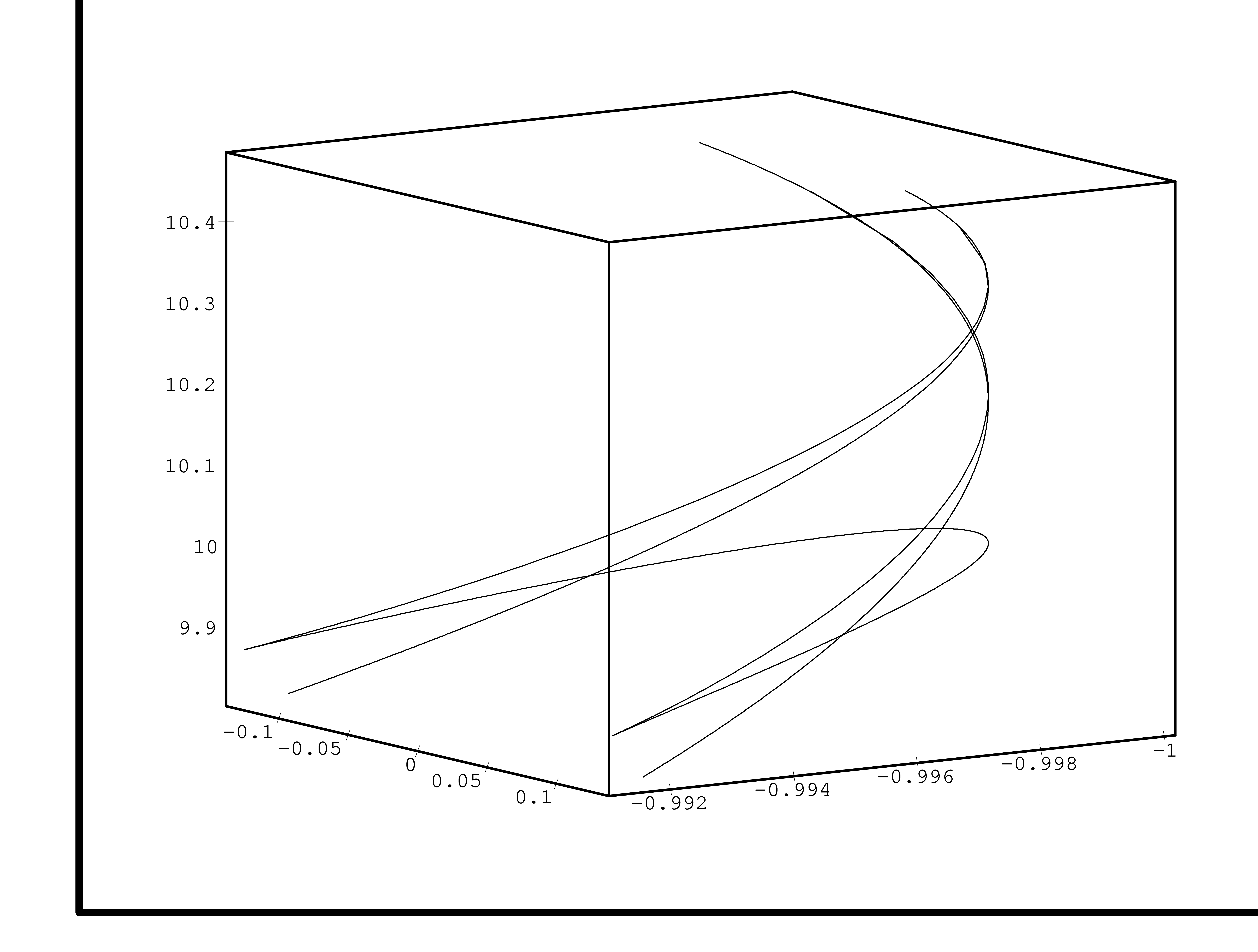,height=3in,width=2.5in}}}
\caption{ Light cone cut for (a) $r_0=10$ and (b) $r_0=5$ }
\end{figure}

As we can see both graphically and analytically, $\displaystyle{\frac{du}{d\phi}}$ is well defined and finite at the caustic points. These caustics are called {\it ``cusps''}. It is rather surprising that, to first order, the singularity structure  around $\theta_c$ and $\theta_0$ is similar. The behaviours of $\displaystyle{\frac{d^2u}{d\phi^2}}$ and  $\displaystyle{\frac{d^3u}{d\phi^3}}$ however, are completely different. Whereas at  $\theta_0$ the second and third derivatives blow up, at $\theta_c$ they are distributional. 

\section{Conclusions}

We close this work with several comments:
\begin{itemize} 

\item The cut function $Z$ only gives a local description of a light
cone cut. Using the available formalism to describe characteristic
surfaces\cite{Arnold} one can show that a light cone cut will have self
intersections (thus $Z$ will not be unique) and singular points where
$\p^2 Z$ and $\p^3 Z$ diverge. It is thus desirable to find alternative
frameworks to describe the cut near singular points and/or globally on
$\scrip$. One of these alternative approaches is given by the method of
generating families of Legendre submanifolds\cite{Arnold} which might
lead to a global description of the light cone cuts.

\item  Likewise, the null hypersurface $Z=const.$ starts as a smooth surface near $\scrip$ but later develops conjugate points and caustics after the surface enters a region with $T_{ab} \neq 0$. A similar approach using generating families might give a global description of these null surfaces. In particular, there is a need to find a version of the metricity condition that will remain smooth around caustic points.

\item We mention again that an interesting and open problem is to study the geometrical meaning of the metricity conditions in four dimensions. We conjecture that they are a torsion free condition on the Cartan connection.

\item The most general singularity of a light cone cut in 3-D can be found on the explicit example presented above. In addition, we find another singular behaviour that is not consider in the literature since the Legendre submanifold is assumed to be at least $C^2$. It appears that by relaxing differentiability conditions on the Legendre submanifold one should be able to prove generic results for this new type of singularity.

\end{itemize} 
\subsection*{Acknowledgments}
 
This research has been partially supported by AIT, CONICET, CONICOR, and UNC. We thank Paul Tod for enlightening discussions.

\appendix
\section{Conformal Einstein metrics and the Bach equation.}

In this Appendix  we derive the field equations for metrics conformally related to Einstein space times. This derivation will be restricted to three and four dimensions.   Let $\hat M$ be a  n-dimensional manifold (n=3 or 4) and $\hat g_{ab}$ a metric with arbitrary signature. Note that we are not following the same conventions as in the main body of the paper where the physical metric is ``unhatted''. This is done to simplify the notation for the conformal metrics and their related tensors. 

Assume that the metric $\hat g_{ab}$ satisfies the Einstein's equation,  i. e. 
$$
\hat R_{ab}-\frac{\hat R}{2}\hat g_{ab}=T_{ab}
$$
where $T_{ab}$ is the stress-energy tensor, and $\hat R_{ab}$, $\hat R$, the Ricci tensor and the Ricci scalar of $\hat g_{ab}$ respectively.  Introducing 
$$
\hat S_{ab}=\frac{1}{n-2}\left (\hat R_{ab} -\frac{\hat R}{2(n-1)}\hat g_{ab}\right),
$$
we can rewrite the equations as
$$
\hat  S_{ab}-\hat S\hat g_{ab}=\frac{1}{(n-2)}T_{ab}.
$$

We shall express this equation in terms of a  conformal metric $g_{ab}$ which is obtained from $\hat g_{ab}$ by rescaling with the conformal factor $\Omega$.  (Note that this conformal factor is not the conformal factor for an asymptotically  flat spacetime as the one given in section IV.)

\begin{equation}
\label{conf}
g_{ab}=\Omega^2\hat g_{ab}
\quad \mbox{or equivalently}\quad 
\hat g^{ab}=\Omega^2 g^{ab}.
\end{equation}

We now write the transformation law of various tensor fields under this conformal rescaling. Let us first decompose the Riemann curvature tensor of $g_{ab}$ according to

$$
R_{abcd}=C_{abcd}+2 (g_{a[c}S_{d]b}-g_{b[c}S_{d]a}) ;
$$
here we assume that $n\geq 3$. In terms of $S_{ab}$ the Bianchi identities can be put as
\begin{equation}
\label{B1}
\nabla^aS_{ab}=\nabla_b S
\end{equation}
 and
\begin{equation}
\label{B2}
\nabla^dC_{abcd}=-2(n-3)\nabla_{[a} S_{b]c}. 
\end{equation}
 Considering the conformal transformation given by (\ref{conf}), 
 the  tensor $\hat S_{ab}$ transforms

\be
\hat S_{ab}\Omega=S_{ab}\Omega+\nabla_a\nabla_b \Omega- \frac{\Omega^{-1}}{2}g_{ab}\nabla_c\Omega\nabla^c\Omega. \label{conft}
\ee
Then the conformal Einsteins equation becomes

 \be
\hat S_{ab}\Omega=S_{ab}\Omega+\nabla_a\nabla_b \Omega- \frac{\Omega^{-1}}{2}g_{ab}\nabla_c\Omega\nabla^c\Omega, \label{Einscon}
\ee
where
$$
\hat S_{ab}=\frac{1}{n-2}\left(T_{ab}-\frac{T}{n-1} g_{ab}\right)
$$
with $T=T_{ab} g^{ab}$. Since the stress-energy tensor satisfies $\hat \nabla^a T_{ab}=0$, for later use, we calculate 
\begin{equation}
\label{divT}
\Omega\nabla^a T_{ab}=(n-2)\nabla^a\Omega T_{ab}-\nabla_b\Omega T.
\end{equation}
 We now ask what conditions should be imposed on $g_{ab}$ so that a
solution of (\ref{Einscon}) exists.  If we consider eq. (\ref{Einscon})
as a second order differential equation for $\Omega$ it is then clear
that, for a non trivial solution to exist, integrability conditions
must be imposed  on the metric $g_{ab}$. Applying $\nabla_c$ to  (\ref{Einscon}), and antisymmetrizing, we get 
\be
\label{derEins0}
\nabla_{[c}\hat S_{a]b}\Omega+\nabla_{[c}\Omega\hat S_{a]b}=\nabla_{[c}S_{a]b}\Omega+\nabla_{[c}\Omega S_{a]b}+
\nabla_{[c}\nabla_{a]}\nabla_b\Omega-\frac{g_{b[a}\nabla_{c]}}{2}(\Omega^{-1} \nabla_d\Omega\nabla^d\Omega).
\ee
Contracting equation (\ref{Einscon}) with $\nabla^a\Omega$ we find
$$
\nabla^a\Omega\hat S_{ab}-\nabla^a\Omega S_{ab}=\frac{\nabla_b}{2}(\Omega^{-1} \nabla_d\Omega\nabla^d\Omega),
$$
and replacing the last term of (\ref{derEins0}) by this result,  yields 

\be
\label{derEins}
\nabla_{[c}\hat S_{a]b}\Omega+\nabla_{[c}\Omega \hat S_{a]b}- g_{b[c}\nabla^d\Omega\hat S_{a]d}=\nabla_{[c}S_{a]b}\Omega+\nabla_{[c}\Omega S_{a]b}+
\nabla_{[c}\nabla_{a]}\nabla_b\Omega-g_{b[c}\nabla^d\Omega S_{a]d}.
\ee
Finally, using the definition of the Riemann tensor, we calculate
$$  
 \nabla_{[c}\nabla_{a]}\nabla_b\Omega= \frac{1}{2}C_{cabd}\nabla^d\Omega+ g_{b[c}S_{a]d}\nabla^d\Omega- \nabla_{[c}\Omega S_{a]b},
$$
 and  equation (\ref{derEins}) becomes  
\be
\label{Einscon1}
2(\nabla_{[a}\hat S_{b]c}\Omega+\nabla_{[a}\Omega \hat S_{b]c}- g_{c[a}\nabla^d\Omega\hat S_{b]d})=2\Omega\nabla_{[a}S_{b]c}+C_{abcd}\nabla^d\Omega
\ee
Now we consider  the  case where the stress-energy tensor is pure trace, i.e.  $$
T _{ab}=-\Lambda\hat g_{ab},
$$
where $\Lambda$ is the cosmological constant. This gives in terms of $g_{ab}$ 
$$
\hat S_{ab}=\frac{\Lambda\Omega^{-2}}{(n-2)(n-1)}g_{ab}. 
$$
Inserting this in the l.h.s. of (\ref{Einscon1}), we have

$$
\nabla_{[a}\hat S_{b]c}\Omega+\nabla_{[a}\Omega \hat S_{b]c}- g_{c[a}\nabla^d\Omega\hat S_{b]d}=\frac{\Lambda\Omega^{-2}}{(n-2)(n-1)}\left(-2\nabla_{[a}\Omega g_{b]c}+\nabla_{[a}\Omega g_{b]c}- g_{c[a}\nabla_{b]}\Omega\right)=0. 
$$
In dimension $n=3$, using the fact that  $C_{abcd}=0$,  equation (\ref{Einscon1}) becomes
\be
\label{bach3}
2\nabla_{[a}S_{b]c}=0.
\ee
It is easy to prove that the tensor $\nabla_{[a}S_{b]c}$ is conformally invariant since
\bea
\label{confbach}
\Omega\nabla_{[a}S_{b]c}&=& \nabla_{[a}\hat S_{b]c}\Omega+\nabla_{[a}\Omega \hat S_{b]c}- g_{c[a}\nabla^d\Omega\hat S_{b]d}\nn\\
&=&\Omega(\hat \nabla_{[a}\hat S_{b]c}-C^l{}_{c[a}S_{b]l})+\nabla_{[a}\Omega \hat S_{b]c}- g_{c[a}\nabla^d\Omega\hat S_{b]d}\nn\\
&=&\Omega\hat \nabla_{[a}\hat S_{b]c},
\eea
where
$$
C^l{}_{ac}=\Omega^{-1} (2\delta^l_{(a }\nabla_{c)}\Omega-g_{ac}\nabla^l\Omega).
$$

In order to obtain the integrability condition for $n>3$,  we use Bianchi identity (\ref{B2}) and equation (\ref{Einscon1}) becomes 
\be
\label{Einscon2}
C_{abcd}\nabla^d\omega-\frac{1}{n-3}\nabla^dC_{abcd}=0,
\ee
where for simplicity in the calculations we have defined $w= \log \Omega $. 
Applying  $\nabla^a$ to (\ref{Einscon2}) we find

$$
\nabla^a\nabla^d\omega C_{abcd} +\nabla^aC_{abcd}\nabla^d\omega-\frac{1}{n-3}\nabla^a\nabla^d C_{abcd}=0
$$
and using  (\ref{Einscon}) and (\ref{Einscon2}), we get

\be 
\label{Einscon3}
\frac{1 }{n-2} R^{ad} C_{abcd}+\frac{1}{n-3}\nabla^a\nabla^dC_{abcd}- \frac{n-4}{n-3}\nabla^d\omega\nabla^aC_{abcd}=0.
\ee
Clearly if $n=4$ we have  the corresponding Bach equation \cite{conf} 

\be
\label{bach4}
B_{bc}:=\frac{R^{ad}}{2} C_{abcd}+\nabla^a\nabla^dC_{abcd}=0.
\ee

\section{ The Bach equation with stress-energy tensor}
In order to handle a more general stress-energy tensor than the pure trace, we may suppose that the stress-energy tensor is given and that it does not  depend on $\Omega$.  

We rewrite equation (\ref{Einscon1})  in terms of the stress-energy tensor and after using (\ref{divT}) we get
\begin{equation}
\label{Einscon5}
2\nabla_{[a}S_{b]c}+C_{abcd}\nabla^d\omega=t_{abc},
\end{equation}
where
$$
t_{abc}=\frac{2}{n-2}\bigg(\nabla_{[a}T_{b]c}-\frac{g_{c[a}\nabla^dT_{b]d}}{n-2}-\frac{\nabla_{[a}T g_{b]c}}{n-1}+\nabla_{[a}\omega T_{b]c}- \frac{(n-3)\;  T}{(n-2)(n-1)}\nabla_{[a}\omega g_{b]c}\bigg).
$$
We shall concentrate our attention to two cases, when the dimensions are $n=3$ 
and $n=4$. In the first case, equation (\ref{Einscon5}) becomes

\begin{equation}
\label{Einscon7}
B_{cab}=2\left(\nabla_{[a} T_{b]c}- g_{c[a}\nabla^d T_{b]d} -\frac{\nabla_{[a} T g_{b]c}}{2} +\nabla_{[a}\omega T_{b]c}\right).
\end{equation}
Contracting this equation with $ T_{bc}$ and  using twice (\ref{divT})  , we get

\begin{equation}
\label{domega}
( T_{bc} T^{bc}-  T^2)\nabla_a \omega=  T^{bc}B_{cab}- 2  T^{bc}
\left(\nabla_{[a} T_{b]c}- g_{c[a}\nabla^d T_{b]d} -\frac{\nabla_{[a} T g_{b]c}}{2}\right)+\nabla^b T_{bc} T_a{}^c+ T\nabla^c T_{ca}.
\end{equation}
Demanding  that $ T_{bc} T^{bc}-  T^2\neq 0$ and inserting (\ref{domega}) in (\ref{Einscon7}), we obtain

\begin{eqnarray}
\label{intcon4}
( T_{de} T^{de}-  T^2) B_{cab}&=&2\left ( T_{de} T^{de}-  T^2\right)\left(\nabla_{[a} T_{b]c}- g_{c[a}\nabla^d T_{b]d} -\frac{\nabla_{[a} T g_{b]c}}{2}\right) + T_{bc} T^{de}B_{ead}-\nonumber\\
&&- T_{ac} T^{de}B_{ebd}- 2  T^{de}
 T_{bc}\left(\nabla_{[a} T_{d]e}- g_{e[a}\nabla^l T_{d]l} -\frac{\nabla_{[a} T g_{d]e}}{2}\right)+\nonumber\\
&&+2  T^{de}
 T_{ac}\left(\nabla_{[b} T_{d]e}- g_{e[b}\nabla^l T_{d]l} -\frac{\nabla_{[b} T g_{d]e}}{2}\right)+\nonumber\\
&&+ T_{bc}\left(\nabla^d T_{de} T_a{}^e+ T\nabla^e T_{ea}\right)- T_{ac}\left(\nabla^d T_{de} T_b{}^e- T\nabla^e T_{eb}\right).
\end{eqnarray}
When $n=4$ equation (\ref{Einscon5}) becomes

\begin{equation}
\label{Einscon8}
C_{abcd}\nabla^d\omega-\nabla^dC_{abcd}=t_{abc},
\end{equation}
where 
\begin{equation}
\label{tabc}
t_{abc}=\nabla_{[a} T_{b]c}-\frac{g_{c[a}\nabla^d T_{b]d}}{2}-\frac{\nabla_{[a} T g_{b]c}}{3}+\nabla_{[a}\omega T_{b]c}- \frac{ T}{6}\nabla_{[a}\omega g_{b]c}.
\end{equation}
As before, applying $\nabla^a$ and  using Einsteins equations we can write

\begin{equation}
\label{incond8}
2B_{ab}= T^{ad}C_{abcd}-2\left(t_{acb}\nabla^a\omega+\nabla^at_{abc}\right).
\end{equation}
A simple calculation gives
\begin{eqnarray*}
2\left(t_{acb}\nabla^a\omega+\nabla^at_{abc}\right)&=&2\nabla^a\bigg(\nabla_{[a} T_{b]c}-\frac{g_{c[a}\nabla^d T_{b]d}}{2}-\frac{\nabla_{[a} T g_{b]c}}{3}\bigg)-\frac{1}{2} T_{ac} \left( T^a{}_{b}-\frac{ T}{6}g^a{}_{b}\right)+\\
&&+\frac{1}{2} T_{ac}R^a{}_b- T\nabla_c\omega\nabla_b\omega+\frac{3}{2}\nabla^a\omega\nabla_a\omega\left( T_{bc}-\frac{ T}{6}g_{bc}\right)+\\
&&+2\nabla^a\omega\nabla_a T_{bc}-\frac{1}{2}\nabla^a\omega\nabla_a T g_{bc}+\frac{1}{2}\nabla_b\omega\nabla_c T- \nabla^a\omega\nabla_c T_{ab}+\\
&&+\frac{1}{2}g_{bc}\nabla^a\omega\nabla^d T_{ad}-\frac{1}{2}\nabla_c\omega\nabla^a T_{ab}-\nabla^a T_{ac}\nabla_b\omega.
\end{eqnarray*}
Contracting (\ref{Einscon8}) with $ C^{abc}{}_l$ and  using twice (\ref{divT}), we get
\begin{eqnarray}
\label{domega4}
(3C_{abcd}C^{abcd}-6 T_{bc} T^{bc}+2 T^2)\nabla_a \omega&=& 12\nabla_d(C^{den}{}_a T_{en})+3\nabla_a(  T^{de} T_{de})- 6\nabla_d(  T_{ae} T^{de})-\nonumber\\
&&-3\nabla^d T_{de} T^{e}{}_a+5 T\nabla^d T_{da}-\nabla_a T^2+2T^{e}{}_a\nabla_e T+\nonumber\\
&&+12 \nabla^lC_{denl} C^{den}{}_a. 
 \end{eqnarray}
Demanding  that $3C_{abcd}C^{abcd}-6 T_{bc} T^{bc}+2 T^2\neq 0$ and inserting (\ref{domega4}) in (\ref{incond8}), we obtain a similar result than for dimension 3.


\end{document}